\documentstyle[twocolumn,prb,eqsecnum,aps,epsf]{revtex}
\begin{document}
\tighten
\draft

\title{\centerline{The incommensurate ground state}
       \centerline{of double-layer quantum Hall systems}}

\author{
C.~B.~Hanna$^{(a)}$, A.~H.~MacDonald$^{(b)}$, and S.~M.~Girvin$^{(c)}$
}
\address{$^{(a)}$Department of Physics, Boise State University,
Boise, Idaho~~83725\\
$^{(b)}$Department of Physics, University of Texas,
Dallas, Texas~~78712\\
$^{(c)}$Department of Physics, Indiana University,
Bloomington, Indiana~~47405\\}

\date{\today}

\maketitle

\begin{abstract}

Double-layer quantum Hall systems possess interlayer phase coherence
at sufficiently small layer separations, even without
interlayer tunneling.
When interlayer tunneling is present,
application of a sufficiently strong in-plane magnetic field
$B_\parallel > B_{\rm c}$ drives a commensurate-incommensurate (CI)
transition to an incommensurate soliton-lattice (SL) state.
We calculate the Hartree-Fock ground-state energy of the SL state
for all values of $B_\parallel$ within a gradient approximation,
and use it to obtain the anisotropic SL stiffness,
the Kosterlitz-Thouless melting temperature for the SL,
and the SL magnetization.
The in-plane differential magnetic susceptibility diverges as
$|B_\parallel - B_{\rm c}|^{-1}$ when the CI transition is
approached from the SL state.

\end{abstract}

%\pacs{PACS numbers: 64.70.Rh, 71.10.Pm, 71.45.Gm,
%                    73.21.Fg, 73.43.Cd, 73.43.Nq}


\section{Introduction}
\label{sec:intro}

At sufficiently low particle densities and small layer separations,
double-layer quantum Hall (2LQH) systems at total filling factor unity
($\nu_T=1$)
can be described as broken-symmetry states\cite{wenzee} that possess interlayer
phase coherence, even in the absence of interlayer
tunneling.\cite{daspin}
The 2LQH system can be mapped to an equivalent spin-1/2
system by equating ``up'' (``down'') pseudospins with electrons
in the upper (lower) layer.\cite{mpb,yangprl,moonbigprb}
(The electrons are assumed to have their real spins polarized.)
The exchange energy between
the electrons produces a pseudospin stiffness $\rho_{\rm s}$
that seeks to keep the pseudospins aligned locally.
At finite layer separation $d$,
the direct (Hartree) energy produces a local capacitive charging
energy that is minimized when the two layers have equal electron density.
Thus the expectation value of the $z$ component of the pseudospin
vanishes and the pseudospin system has an ``easy-plane'' anisotropy
that gives the itinerant ferromagnet an $XY$
symmetry\cite{yangprl,moonbigprb}
(in the absence of interlayer tunneling).
The expectation value of a pseudospin at location ${\bf r}$ can
therefore be specified by its angle $\theta({\bf r})$ in the $xy$ plane.

In the absence of interlayer tunneling,
the 2LQH system picks out a particular global value of $\theta$
in the ground state,
spontaneously breaking the $U(1)$ symmetry of the $XY$ ferromagnet.
This gives rise to a linearly dispersing Goldstone mode at long
wavelengths.\cite{yangprl,moonbigprb}
Recent measurements of the zero-bias tunneling conductance
in 2LQH systems show a huge resonant enhancement when
interlayer coherence is present.\cite{spielman}
This enhancement is related to the Goldstone mode of the
broken $U(1)$ symmetry, and it has been proposed that the dispersion
of the Goldstone mode can be observed in tunneling conductance measurements
by applying a weak parallel magnetic field.\cite{tuntheory}
The $XY$ pseudomagnet also possesses
vortex excitations called ``merons'';
unlike those in an ordinary ferromagnet,
these vortices are electrically charged, and the lowest-energy
charged excitations of the system consist of vorticity-neutral
meron pairs.\cite{yangprl,moonbigprb}
There is also a finite-temperature Kosterlitz-Thouless (KT)
phase transition due to the $XY$ symmetry and the finite
pseudospin stiffness of the ferromagnet.

In the presence of interlayer tunneling, the $U(1)$ invariance
associated with conservation of the charge difference
between the two layers is lost.
The finite interlayer tunneling $t$ of the electrons acts as an
effective Zeeman pseudofield of magnitude $2t$ along the
pseudospin $x$ axis
and aligns the pseudospins, so that $\theta=0$ in the ground state.
The Goldstone mode disappears, and the collective mode becomes gapped.
In the presence of interlayer tunneling, merons of opposite vorticity
are bound together by a ``string'' that produces a linear
attraction between the merons, eliminating the finite-temperature
KT transition.\cite{yangprl,yangbigprb}

Murphy {\it et al.} have investigated the effect of an in-plane magnetic
field $B_\parallel$ on 2LQH systems.\cite{murphy}
By varying $B_\parallel$ and studying the energy gap obtained from
activation energy measurements of the longitudinal resistivity,
they find evidence for a phase transition between two competing
QH ground states at a critical value $B_\parallel = B_{\rm c}$.
These two ground states are understood in the pseudospin language
as being due to a competition between the tunneling energy $t$
and the exchange energy $\rho_{\rm s}$.\cite{yangprl,moonbigprb}
Read has presented an appealing analysis of charged (meron pair)
excitations in this system, focusing on the value of the charge gap
near the commensurate-incommensurate transition.\cite{read}

The presence of $B_\parallel$ periodically shifts the phase of
the tunneling matrix elements, resulting in an effective Zeeman
field for the pseudospins that rotates along the planar direction
perpendicular to ${\bf B_\parallel}$,
 with a wavelength $\phi_0/B_\parallel d$,
which is the distance required to contain one flux quantum
$\phi_0=h/e$ between the layers.
The net result is that for gradual variations of the pseudospin texture
(gradual on the scale of the magnetic length
$\ell\equiv\sqrt{\hbar/eB_\perp}$, where $B_\perp$ is the strength
of the magnetic field normal to the plane),
the energy of the $XY$ pseudospin system has the
Pokrovsky-Talapov (PT) form,\cite{pokrovsky,bak,dennijs}
\begin{equation}
\label{eq:ept}
{\cal E} = \int d^2r \left[
\frac{1}{2} \rho_{\rm s} \left({\mathbf \nabla}\theta\right)^2 -
\frac{t}{2\pi\ell^2} \cos (\theta + {\bf Q\cdot r})
\right] ,
\end{equation}
up to a constant, where
${\bf Q} \equiv (2\pi d/\phi_0) {\bf B_\parallel\!\times\!\hat{z}}$
defines the parallel magnetic-field wave vector,
\begin{equation}
\label{eq:tnu}
t = t_0 e^{-Q^2\ell^2/4} \sqrt{1-m_z^2}
\end{equation}
is the tunneling energy
(where $t_0$ is the tunneling energy when $Q=0$),\cite{hu} and
\begin{equation}
\label{eq:rhonu}
\rho_{\rm s} = (1-m_z^2) \rho_E
\end{equation}
is the pseudospin stiffness
within the Hartree-Fock Approximation (HFA).
Here $m_z \equiv \nu_1-\nu_2$ is the layer imbalance, and
\begin{equation}
\rho_E = \frac{e^2}{4\pi\epsilon\ell} \frac{1}{16\pi}
\int_0^\infty dx x^2 e^{-x^2/2} e^{-xd/\ell}
\end{equation}
is the interlayer exchange stiffness when the layers are
balanced: $\nu_1=\nu_2=1/2$, or $m_z=0$.
The layer separation is $d$, and
$\nu_j$ is the filling factor of layer $j$.

For small $Q$ (small $B_\parallel$), the ground-state energy
is minimized by having the pseudospins point in the direction
of the local (rotating) pseudospin Zeeman field, so that
$\theta({\bf r})=-{\bf Q}\cdot {\bf r}$.
This is the commensurate ground state, and it
minimizes the pseudospin Zeeman (tunneling) energy.
However, it does so at the expense of the exchange energy,
since neighboring pseudospins are no longer parallel.
In the limit of large $Q$, the cost in exchange energy
for the pseudospins to align with the rapidly rotating Zeeman field
is prohibitive, and the pseudospins become (nearly) uniformly
polarized (constant $\theta$), just as if $t\rightarrow 0$.
The state with uniformly polarized pseudospin is
the large-$Q$ limit of the incommensurate state.
It turns out that for all finite $B_\parallel > B_{\rm c}$,
the translational symmetry of the pseudospin polarization is broken,
and a soliton-lattice (SL) state is obtained in the incommensurate phase.

The SL phase of the PT model can also undergo a separate
finite-temperature dislocation-mediated KT transition that restores
the translational symmetry.\cite{dennijs}
This work focuses on calculating the ground-state properties of
the SL state, for all $B_\parallel > B_{\rm c}$.
Interestingly, it is not necessary to determine
$\theta({\bf r})$ in order to calculate the total energy of
the system.\cite{bak,dennijs,perring}
{}From the total energy, we calculate thermodynamic quantities such as
the SL stiffnesses, extending the results of Ref.~\onlinecite{read}
for the stiffnesses and the KT temperature to all $B_\parallel$.
We also calculate the SL contribution to the magnetization and
susceptibility, and discuss some possibilities of measuring these
quantities experimentally.

The plan of this paper is as follows.
In Sec.~\ref{sec:singsols}, we discuss the single-soliton solutions
that follow from the equation of motion obtained from the PT energy.
For $Q>Q_{\rm c}$ the solitons proliferate; in Sec.~\ref{sec:solat},
the interaction between soliton lines is discussed, and the separation
between solitons as a function of $Q/Q_{\rm c}$ is derived.
In Sec.~\ref{sec:stiff}, the compressional and shear elastic constants
are analyzed, and an estimate is made of the Kosterlitz-Thouless
temperature for melting the soliton lattice, as a function of $Q$.
The interlayer phase coherent 2LQH state has a diamagnetic response
to an applied in-plane magnetic field;\cite{ep2ds12}
Section~\ref{sec:mag} gives a calculation of the in-plane magnetization
due to the 2LQH state, as a function of $Q$.
We conclude with a summary of our results.

\section{Single Solitons}
\label{sec:singsols}

When $Q>Q_{\rm c}$, it is convenient to define
$\tilde{\theta}({\bf r}) \equiv \theta({\bf r}) + {\bf Q\cdot r}$,
so that Eq.~(\ref{eq:ept}) becomes
\begin{equation}
\label{eq:tp2}
{\cal E} = \int d^2r \left [
\frac{1}{2}\rho_{\rm s}
\left({\mathbf \nabla}\tilde{\theta} - {\bf Q}\right)^2 -
\frac{t}{2\pi\ell^2} \cos \tilde{\theta} \right ] .
\end{equation}
Minimizing ${\cal E}$ with respect to variations in
$\tilde{\theta}$ gives the sine-Gordon equation:
\begin{equation}
\label{eq:sg1}
\nabla^2 \tilde{\theta} = \frac{1}{\xi^2} \sin \tilde{\theta} ,
\end{equation}
where $\xi/\ell = \sqrt{2\pi\rho_{\rm s}/t}$.
We shall give numerical values for our results
for a hypothetical ``typical'' GaAs
(effective mass $m^*\approx 0.07m_e$ and relative dielectric constant
$\epsilon_r\approx 13$) 2LQH sample, which for the sake
of definiteness we assign the following sample parameters:
total particle areal density $n_T=1.0\times 10^{11}~\rm{cm}^{-2}$,
layer (midwell to midwell) separation $d=20$~nm,
and tunneling energy $t_0=0.1$~meV.
Such a sample would have $\ell\approx 12.6$~nm, $d/\ell\approx 1.6$,
$\hbar\omega_{\rm c}\approx 6.9$~meV for $\nu_{\rm T}=1$, and
$e^2/4\pi\epsilon\ell\approx 8.8$~meV.
In the HFA,
$\rho_{\rm E}\approx 0.03$~meV and $\xi\approx 17$~nm.

\begin{figure}[t]
\epsfxsize4.0in
\centerline{\epsffile{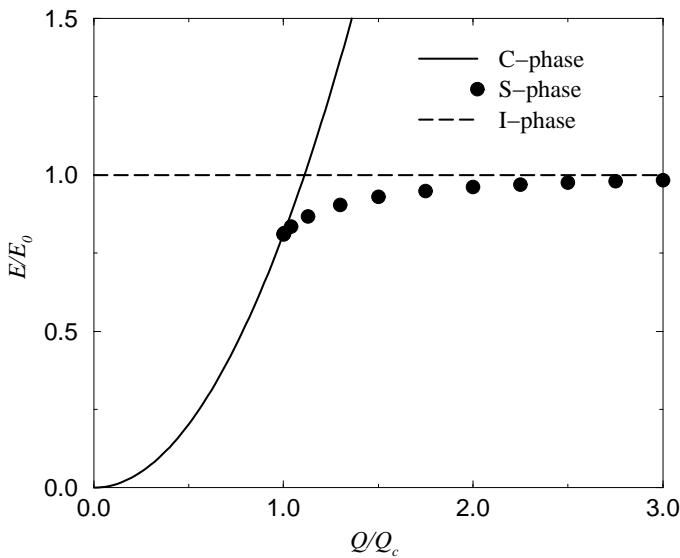}}
\caption{Energy of the commensurate (C), constant-$\theta$ incommensurate(I),
and incommensurate SL (S) phases, versus the parallel magnetic-field
wave vector $Q$.
For $Q \ge Q_{\rm c}$, the SL phase (dotted) has the lowest energy.}
\label{hannafig1}
\end{figure}

The commensurate state minimizes the tunneling energy by having
$\tilde{\theta}({\bf r}) = 0$,
so that the phase angle $\theta({\bf r}) = {\bf Q \cdot r}$
follows the tumbling Zeeman pseudofield.  The energy per area
of this state is $\rho_{\rm s} Q^2/2- t/2\pi\ell^2$.
In the limit of large $Q$, the incommensurate state with constant
$\theta$ has a lower energy per area, equal to zero.
These energies are plotted as the solid and dashed lines in
Fig.~\ref{hannafig1}.
We therefore estimate that there is a phase transition near the point
where the commensurate-state and the constant-$\theta$
incommensurate-state energies are equal, at $Q\xi\approx\sqrt{2}$.
It turns out, however, that the incommensurate state lowers its
energy by breaking translation invariance, so that at finite $t$ and
$Q>Q_c$, the value of $\tilde{\theta}$ depends on position.

Equation~(\ref{eq:sg1}) possesses soliton solutions.
To see this, let us seek solutions of the form
\begin{equation}
\label{eq:thform}
\tilde{\theta}({\bf r}) =
\tilde{\theta}\left[{\bf\hat{e}}_1 \cdot ({\bf r}-{\bf r}_0)\right] ,
\end{equation}
where ${\bf\hat{e}}_1$ could be any unit vector in the $xy$ plane.
Then Eq.~(\ref{eq:sg1}) becomes
\begin{equation}
\label{eq:sg2}
\partial^2_1 \tilde{\theta} = \frac{1}{\xi^2} \sin \tilde{\theta} .
\end{equation}
Note that this equation is equivalent to the equation of motion for
a pendulum of length $l$, \mbox{$\partial^2_t\phi = -(g/l) \sin\phi$},
if we replace \mbox{$\tilde{\theta} \rightarrow \phi - \pi$},
\mbox{${\bf\hat{e}}_1 \cdot {\bf r} \rightarrow t$} (time),
\mbox{$\rho_{\rm s} \rightarrow l$}, and 
(tunneling amplitude) \mbox{$t/2\pi\ell^2 \rightarrow g$}.
This analogy is very useful in finding the soliton solutions for the
PT model.
In particular, the pendulum can make full circles in a given direction.
This corresponds to the SL state in the 2LQH system.

In analogy with the pendulum problem, we may
define a conserved quantity analogous to the total
(kinetic plus potential) energy of a pendulum:
\begin{equation}
\label{eq:q2}
2 c^2 \equiv
\frac{1}{2} \xi^2 \left ( \partial_1 \tilde{\theta} \right )^2 -
(1 - \cos \tilde{\theta}) .
\end{equation}
Differentiating Eq.~(\ref{eq:q2}) with respect to
${\bf\hat{e}}_1\cdot {\bf r}$ and using Eq.~(\ref{eq:sg2})
shows that $\partial_1 c = 0$, so that $c$ is a constant of the motion.
Defining $\beta=\tilde{\theta}/2$ then leads to the equation
\begin{equation}
\label{eq:beta}
\partial_1 \beta = \pm \frac{1}{\xi}
\sqrt{c^2 +  \sin^2 \beta} .
\end{equation}

\begin{figure}[b]
\epsfxsize4.0in
\centerline{\epsffile{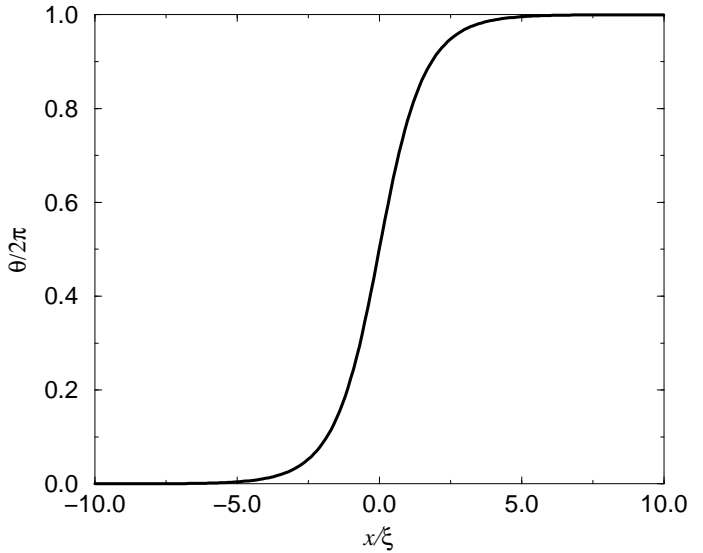}}
\caption{At $Q=Q_{\rm c}$, the system admits a single soliton, in
which $\tilde{\theta}({\bf r})$ twists by $2\pi$ over a distance
$\xi$.}
\label{hannafig2}
\end{figure}

It is straightforward to solve Eq.~(\ref{eq:beta}) when $c=0$
by writing $f = \tan(\beta/2)$, so that
$\partial_1 f = \pm f/\xi$, giving
$\tilde{\theta} = \tilde{\theta}_{\rm ss}({\bf r})$, where
\begin{equation}
\label{eq:sol}
\tilde{\theta}_{\rm ss}({\bf r}) = 4 \arctan \left [ 
e^{\pm {\bf\hat{e}}_1\cdot ({\bf r} - {\bf r}_0)/\xi} \right ]
\end{equation}
represents a single soliton in the ${\bf\hat{e}}_1$-direction,
centered at ${\bf\hat{e}}_1\cdot {\bf r}_0$,
with width $\xi$.
This is shown in Fig.~\ref{hannafig2}.
Note that $\tilde{\theta}_{\rm ss}({\bf r})$ changes by $2\pi$
as ${\bf\hat{e}}_1\cdot{\bf r}$ goes from $-\infty$ to $\infty$.
This corresponds to the motion of a pendulum that just barely
completes a full revolution, and whose period goes to infinity.

The energy per length of a single soliton may be computed
by substituting Eq.~(\ref{eq:sol}) into Eq.~(\ref{eq:tp2})
and subtracting the commensurate-state ($\tilde{\theta}=0$) energy,
to obtain
\begin{equation}
\label{eq:esol}
\frac{{\cal E}_{\rm ss}}{L_2} = \rho_{\rm s} \left ( \frac{8}{\xi} \mp
2\pi {\bf\hat{e}}_1 \cdot {\bf Q} \right ) ,
\end{equation}
where $L_2$ is the sample length in the planar direction perpendicular
to ${\bf\hat{e}}_1$.
The lowest (soliton) and highest (anti-soliton) energy solutions occur for
${\bf\hat{e}}_1 = \pm{\bf\hat{Q}}$.
Since solitons in the $-{\bf\hat{Q}}$ directions are equivalent to
anti-solitons in the ${\bf\hat{Q}}$, we shall speak only about solitons
with orientations defined by ${\bf\hat{e}}_1$.
The lowest-energy soliton state has ${\bf\hat{e}}_1 = {\bf\hat{Q}}$,
and its energy per length is
\begin{equation}
\label{eq:essbar}
\frac{\bar{{\cal E}}_{\rm ss}}{L_2} =
2\pi \rho_{\rm s} (Q_{\rm c} - Q) ,
\end{equation}
which goes to zero when $Q = Q_{\rm c}$, where
\begin{equation}
\label{eq:qcdef}
Q_{\rm c} \equiv \frac{4}{\pi\xi} =
\frac{4}{\pi} \sqrt{\frac{t/\rho_{\rm s}}{2\pi\ell^2}} .
\end{equation}

The value of the critical wave vector $Q_c$ will depend on the layer
imbalance $m_z \equiv \nu_1-\nu_2$.
Equations (\ref{eq:tnu}) and (\ref{eq:rhonu}) give
\begin{equation}
\label{eq:qcmz}
Q_c(m_z) = (1-m_z^2)^{-1/4} Q_c(0)
\end{equation}
in the HFA,
where $Q_c(0)$ is the value of $Q_c$ when the layers are balanced
($m_z=0$).
Equation (\ref{eq:qcmz}) implies that the value of $Q_c$ where the CI
transition occurs could be fine-tuned by adjusting the layer imbalance
$m_z$ -- i.e., by adjusting the gate voltages on the 2LQH sample.
Such a procedure might be very useful in investigating the CI transition.

\begin{figure}[t]
\epsfxsize4.0in
\centerline{\epsffile{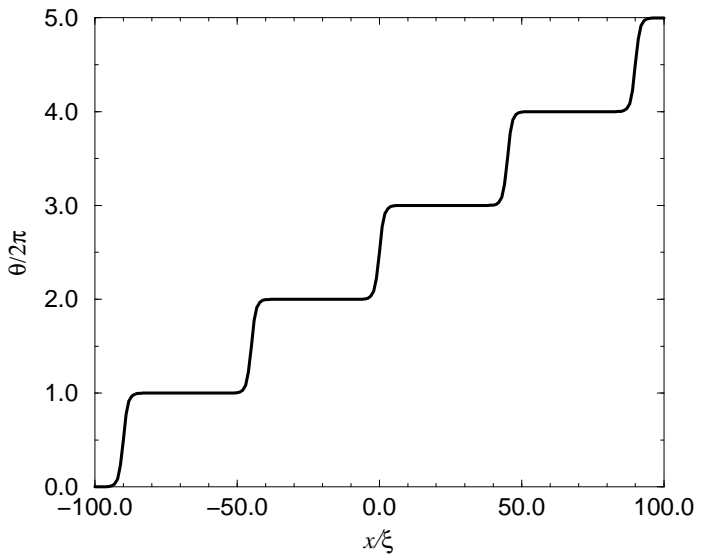}}
\caption{Soliton-lattice for $Q$ slightly larger than $Q_{\rm c}$.}
\label{hannafig3}
\end{figure}

When $Q<Q_c$ and $t>0$, the lowest-energy charged excitations are
finite-length soliton lines with charged meron ends -- i.e., charged vortices
bound by a soliton ``string'' whose tension is given by
Eq.~(\ref{eq:essbar}).
As $Q$ increases, the soliton-line ``string tension'' gets weaker,
so that the Coulomb repulsion of the merons stretches out the
string and lowers the energy of the charged excitation.\cite{daspin}
At $Q=Q_c$ the soliton-line ``string tension'' goes to zero, and
it costs zero energy to make infinitely long soliton lines.
Since the creation energy per length for a soliton
decreases linearly with $Q$ (with $B_\parallel$)
for $Q\ge Q_{\rm c}$, it becomes energetically favorable to
form solitons.
The number of solitons created is determined by the competition between
the (negative) creation energy per soliton versus the repulsive
(positive energy) interactions between solitons.
Note that $Q_{\rm c} < \sqrt{2}/\xi$ (the value of $Q$ at which
the commensurate-state and constant-$\theta$ incommensurate state
energies are equal), so that
for $Q>Q_{\rm c}$ it is energetically favorable to create solitons.
Because the solitons are weakly repulsive, the result is a soliton-lattice
state that we describe below and illustrate in Fig.~\ref{hannafig3}.
An analogous effect occurs in long Josephson junctions, where
application of a magnetic field parallel to two superconducting
planes in close proximity produces $2\pi$ twists in the Josephson
phase and generates a SL state.\cite{lebwohl,fetter}

To summarize, for  $Q<Q_{\rm c}$, we obtain the commensurate phase
in which $\tilde{\theta}({\bf r})=0$.
For $Q=Q_{\rm c}$, we introduce a single soliton,
corresponding to the marginal case of a pendulum that makes exactly
one full revolution and has an infinite period of oscillation.
For $Q>Q_{\rm c}$, we obtain a soliton lattice, corresponding to
a pendulum making complete revolutions in one direction.
It has been argued that the commensurate to incommensurate soliton-lattice
(CI) transition at $Q=Q_{\rm c}$ can be identified with the
phase transition between 2LQH ground states seen by Murphy and
co-workers;\cite{yangprl,murphy}
we therefore make the identification
$Q_{\rm c}=2\pi B_{\rm c}d/\phi_0$.

\section{Soliton Lattice}
\label{sec:solat}

We shall now use Eq.~(\ref{eq:beta}) to determine the SL
spacing $L_{\rm s}$.
We do this by noting that over one period of the soliton lattice,
$\tilde{\theta}$ changes by $2\pi$, so that $\beta$ changes by $\pi$.
We therefore express $L_{\rm s}$ as
\begin{eqnarray}
\label{eq:ls}
\frac{L_{\rm s}}{\xi} &=& \frac{1}{\xi}
\int_{-\pi/2}^{\pi/2} \frac{d\beta}{\partial_1\beta}
\\ \nonumber
&=& 2 \int_0^{\pi/2}\frac{d\beta}{\sqrt{c^2 + \sin^2\beta}}
= 2\eta K(\eta) ,
\end{eqnarray}
where we have used Eq.~(\ref{eq:beta}) and have defined
$\eta \equiv 1/\sqrt{c^2 + 1}$, and where
\begin{eqnarray}
\label{eq:kfunc}
&& K(\eta) \equiv \int_0^{\pi/2}
\frac{d\beta}{\sqrt{1 - \eta^2\sin^2\beta}}
\\ \nonumber
&\rightarrow& \left \lbrace
\begin{array}{ll}
\ln (\frac{4}{\sqrt{1-\eta^2}})
+\frac{1}{4}(1-\eta^2) \left\lbrack
\ln (\frac{4}{\sqrt{1-\eta^2}}) - 1 \right\rbrack ,
& \eta \rightarrow 1 \\
\frac{\pi}{2}
\left ( 1 + \frac{1}{4}\eta^2 +\frac{9}{64}\eta^4 \right ) ,
& \eta \rightarrow 0
\end{array} \right .
\end{eqnarray}
is the complete elliptic integral of the first kind.\cite{gr}
We define the SL wave vector
${\bf Q}_{\rm s} \equiv (2\pi/L_{\rm s}){\bf\hat{e}}_1$, so that
Eq.~(\ref{eq:ls}) may be written in terms of
$Q_{\rm s} \equiv |{\bf Q}_{\rm s}|$ as
\begin{equation}
\label{eq:qsoqc}
\frac{Q_{\rm s}}{Q_{\rm c}} = \frac{(\pi/2)^2}{\eta K(\eta)} .
\end{equation}
Note that $\eta \rightarrow 1$ corresponds to the CI transition,
where $Q_{\rm s} \rightarrow 0$ and $Q \rightarrow Q_{\rm c}$,
whereas $\eta \rightarrow 0$ corresponds to the constant-$\theta$
incommensurate state,
where $Q_{\rm s} \rightarrow Q \rightarrow \infty$.
{}From Eqs. (\ref{eq:qsoqc}) and (\ref{eq:kfunc}), it follows that
\begin{equation}
\label{eq:etaofqs}
\eta \rightarrow \left \lbrace
\begin{array}{lll}
1 - 8e^{-2\pi/Q_{\rm s}\xi} ,
& Q_{\rm s}/Q_{\rm c}\rightarrow 0 \\
\left ( \frac{\pi}{2}\frac{Q_{\rm c}}{Q_{\rm s}} \right )
\left\lbrack 1 -
\frac{1}{4} \left ( \frac{\pi}{2}\frac{Q_{\rm c}}{Q_{\rm s}} \right )^2
\right.
& \\ \qquad \qquad \quad \left.
+ \frac{3}{64} \left ( \frac{\pi}{2}\frac{Q_{\rm c}}{Q_{\rm s}} \right )^4
\right\rbrack ,
& Q_{\rm s}/Q_{\rm c}\rightarrow\infty
\end{array} \right .
\end{equation}

In order to determine the value
$\bar{\bf Q}_{\rm s}$ of the SL wave vector
that minimizes the total energy, we express the energy
per area from Eq.~(\ref{eq:tp2}) as an integral
over $\beta$ [cf. Eq.~(\ref{eq:ls})] and obtain
\begin{eqnarray}
\label{eq:elxly}
&& \frac{{\cal E}}{L_1L_2} = \frac{\rho_{\rm s}}{\xi^2} \left[
\left(\frac{1}{2}Q^2 - {\bf Q}\cdot {\bf Q}_{\rm s}\right) \xi^2
\right.
\\ \nonumber
&& \qquad \qquad
\left. \qquad
+ Q_{\rm c} Q_{\rm s} \xi^2 \frac{E(\eta)}{\eta}
- \left(\frac{2}{\eta^2} - 1\right)
\right] \\ \nonumber
&\rightarrow& \left \lbrace
\begin{array}{llll}
\frac{1}{2}\rho_{\rm s}Q^2 - t/2\pi\ell^2 ,
& Q< Q_{\rm c} \\
\rho_{\rm s} \left[
\frac{1}{2}Q^2 - {\bf Q} \cdot {\bf Q}_{\rm s} \right.
& ~ \\ \left.
\qquad + Q_{\rm c} Q_{\rm s} \left( 1 +
4 e^{-\frac{2\pi}{Q_{\rm s}\xi}} \right) - 1/\xi^2 \right] ,
& \eta \rightarrow 1 \\
\rho_{\rm s} \left [
\frac{1}{2} \left({\bf Q}-{\bf Q}_{\rm s}\right)^2 -
1/\left(2Q_{\rm s}\xi^2\right)^2 \right] ,
& \eta \rightarrow 0
\end{array} \right .
\end{eqnarray}
where
\begin{eqnarray}
\label{eq:efunc}
&& E(\eta) \equiv \int_0^{\pi/2}
d\beta \sqrt{1 - \eta^2\sin^2\beta}
\\ \nonumber
&& \rightarrow \left \lbrace
\begin{array}{cc}
1 + \frac{1}{2}(1-\eta^2)
\left\lbrack \ln (\frac{4}{\sqrt{1-\eta^2}}) -
		  \frac{1}{2} \right\rbrack ,
& \eta \rightarrow 1 \\
\frac{\pi}{2}
\left ( 1 - \frac{1}{4}\eta^2 -\frac{3}{64}\eta^4 \right ) ,
& \eta \rightarrow 0
\end{array} \right .
\end{eqnarray}
is the complete elliptic integral of the second kind,\cite{gr}
and $L_1$ is the sample length in the planar direction parallel
to ${\bf\hat{e}}_1$.
Agreement between Eqs.~(\ref{eq:esol}) and (\ref{eq:elxly})
in the thermodynamic limit is obtained by equating
${\bf Q}_{\rm s} =  (2\pi/L_1) {\bf\hat{e}}_1$,
so that $L_{\rm s} = L_1$ at $Q=Q_{\rm c}$.

All the terms in the $\eta\rightarrow 1$ limit (near the CI transition,
where the solitons are well separated)
of Eq.~(\ref{eq:elxly}) have simple physical interpretations.
The first and last terms,
$\rho_{\rm s} (Q^2/2 - 1/\xi^2)$,
constitute the (commensurate-phase) energy per area of the PT model
in the absence of solitons.
The second and third terms,
\mbox{$\rho_{\rm s} (Q_{\rm c}Q_{\rm s} -
{\bf Q}\cdot {\bf Q}_{\rm s})$}
are just the creation energy per area for the solitons
in terms of interacting soliton lines [see Eq.~(\ref{eq:essbar})].
The fourth term,
$4\rho_{\rm s} Q_{\rm c} Q_{\rm s} \exp(-2\pi/Q_{\rm s}\xi)$,
is the exponentially weak repulsive interaction energy per area
between the solitons.
Because $Q_{\rm s}/2\pi = 1/L_{\rm s}$ is the density of soliton lines,
the fourth term says that near the CI transiton, the interaction
energy per length between two parallel, straight, and infinitely long
solitons lines separated by a distance $L_{\rm s}$ is
\begin{equation}
\label{eq:uly}
\lim_{\eta\rightarrow 1} \frac{U}{L_y} \rightarrow
8\pi \rho_{\rm s} Q_{\rm c} e^{-L_{\rm s}/\xi} .
\end{equation}
Hence for $Q>Q_{\rm c}$ soliton lines proliferate rapidly until
the repulsion between the solitons become significant.
The notion of an effective repulsive interaction between
sine-Gordon solitons was discussed by Perring and Skyrme,\cite{perring}
who obtained the exponentially weak repulsion found above.
The arguments of Ref.~\onlinecite{perring} imply that
when the solitons are close together
($L_{\rm s}/\xi\rightarrow 0$) at large $Q/Q_{\rm c}$,
the repulsive potential energy per length between soliton lines is
$(\pi^3/2) \rho_{\rm s} Q_{\rm c} (\xi/L_s)$.
This latter repulsion is due to boundary condition that $\tilde{\theta}$
must change by $2\pi$ over the small distance $L_{\rm s}$, which
implies a large gradient energy.

The value of ${\bf Q}_{\rm s}$ which minimizes the energy per area
[Eq.~(\ref{eq:elxly})] is found by setting to zero
\begin{equation}
\label{eq:eom}
\frac{\partial}{\partial {\bf Q}_{\rm s}}
\left( \frac{{\cal E}}{L_1L_2} \right)_{\bf Q} =
\rho_{\rm s} \left \lbrack
Q_{\rm c} \frac{E(\eta)}{\eta} {\bf\hat{Q}}_{\rm s} - {\bf Q} ,
\right \rbrack
\end{equation}
where we have used the identity\cite{gr}
\begin{equation}
\label{eq:dedn}
\frac{dE}{d\eta} = \frac{E(\eta) - K(\eta)}{\eta} .
\end{equation}
It is not difficult to show that the second derivative of the
energy per area with respect to $Q_{\rm s}$ is always
positive; thus the extremum calculated above is a minimum.
It follows from Eq.~(\ref{eq:eom}) that the energy is minimized for
${\bf\hat{Q}}_{\rm s} = {\bf\hat{Q}}$, and for $\eta=\bar{\eta}$ such that
\begin{equation}
\label{eq:qoqc}
\frac{Q}{Q_{\rm c}} = \frac{E(\bar{\eta})}{\bar{\eta}} ,
\end{equation}
where $Q \equiv |{\bf Q}|$.

We define the CI closeness parameter
\begin{equation}
\label{eq:epsilon}
\epsilon \equiv Q/Q_{\rm c}-1 ,
\end{equation}
which vanishes at the CI transition;
from Eqs. (\ref{eq:efunc}) and (\ref{eq:qoqc}),
it follows that for $\bar{\eta}\rightarrow 1$,
\begin{equation}
\label{eq:etaofq1}
\epsilon \approx
\left(\frac{1-\bar{\eta}}{2}\right)
\left[\ln\left(\frac{8}{1-\bar{\eta}}\right) + 1\right] ,
\end{equation}
so that asymptotically,
\begin{equation}
\label{eq:etabar}
\bar{\eta} \rightarrow \left \lbrace
\begin{array}{lll}
1 - 2\epsilon/\ln(1/\epsilon) ,
& Q_{\rm s}/Q_{\rm c}\rightarrow 0 \\
\left ( \frac{\pi}{2}\frac{Q_{\rm c}}{Q} \right )
\left\lbrack 1 -
\frac{1}{4} \left ( \frac{\pi}{2}\frac{Q_{\rm c}}{Q} \right )^2
\right.
& \\ \left. \qquad \qquad \quad
+ \frac{5}{64} \left ( \frac{\pi}{2}\frac{Q_{\rm c}}{Q} \right )^4
\right\rbrack ,
& Q_{\rm s}/Q_{\rm c}\rightarrow\infty
\end{array} \right .
\end{equation}

\begin{figure}[b]
\epsfxsize4.0in
\centerline{\epsffile{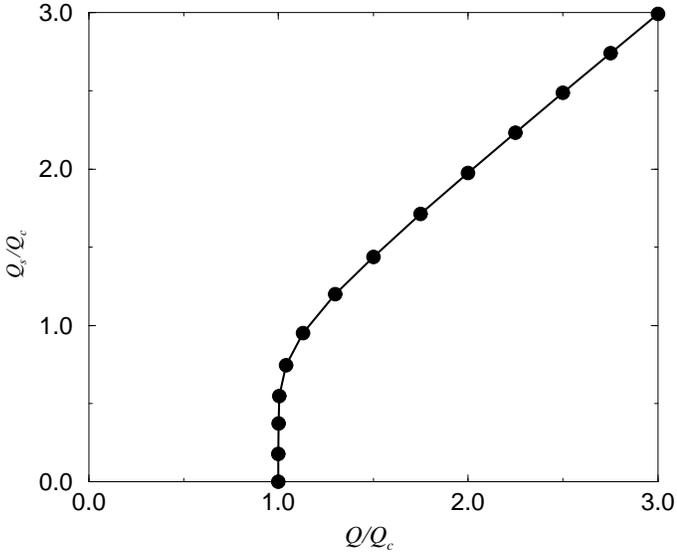}}
\caption{Soliton wave vector $\bar{Q}_{\rm s}$
versus the in-plane magnetic-field wave vector $Q$.
As $Q\rightarrow Q_{\rm c}$,
$\bar{Q}_{\rm s}\xi\sim 2\pi/\ln[Q_{\rm c}/(Q-Q_{\rm c})]$.}
\label{hannafig4}
\end{figure}

Equations~(\ref{eq:qsoqc}) and (\ref{eq:qoqc}) together
allow us to determine the equilibrium SL wave vector
$\bar{\bf Q}_{\rm s}({\bf Q})$ that minimizes the energy,
in terms of the parallel-field wave vector ${\bf Q}$.\cite{bakcorrect}
We have plotted this in Fig.~\ref{hannafig4}.
{}From Eqs. (\ref{eq:qsoqc}), (\ref{eq:qoqc}),
and (\ref{eq:etabar}), it follows that
\begin{eqnarray}
\label{eq:qsofq}
\frac{\bar{Q}_{\rm s}}{Q}
&=& \frac{(\pi/2)^2}{K(\bar{\eta})E(\bar{\eta})}
\\ \nonumber
&\rightarrow& \left \lbrace
\begin{array}{cc}
\left(\pi^2/2\right)/\ln(1/\epsilon) ,
& Q/Q_{\rm c}\rightarrow 1 \\
1 - \frac{1}{32} \left(\frac{\pi}{2}\frac{Q_{\rm c}}{Q}\right)^4 ,
& Q/Q_{\rm c}\rightarrow\infty
\end{array} \right .
\end{eqnarray}
where the $Q\rightarrow Q_c$ limit is true asymptotically.\cite{read,bak,frank}
We note however that, as found by Pokrovsky and Talapov\cite{pokrovsky}
and discussed in Ref.~\onlinecite{fisher2}, the meandering of soliton lines
at finite temperature renormalizes the dependence of the soliton-line
density on the parallel magnetic field, so that
$Q_{\rm s} \propto \sqrt{\epsilon}$
sufficiently close to the CI transition.

The minimized value $\bar{{\cal E}}/L_1L_2$
of the energy per area at $Q_{\rm s}=\bar{Q}_{\rm s}$
is found by using Eqs. (\ref{eq:elxly}) and (\ref{eq:qoqc})
in Eq.~(\ref{eq:efunc}) to obtain
\begin{eqnarray}
\label{eq:emin}
\frac{\bar{{\cal E}}}{L_1L_2} &=&
\frac{1}{2} \rho_{\rm s} Q^2
- \frac{t}{2\pi\ell^2} \left(\frac{2}{\bar{\eta}^2} - 1\right)
\\ \nonumber
&\rightarrow& \left \lbrace
\begin{array}{lll}
\frac{1}{2}\rho_{\rm s}Q^2 - t/2\pi\ell^2 ,
& Q< Q_{\rm c} \\
-\frac{1}{2} \rho_{\rm s} Q_{\rm c}^2 (\pi^2/8-1) ,
& Q/Q_{\rm c} \rightarrow 1 \\
-\frac{t}{2\pi\ell^2} \frac{1}{16}
                      \left(\frac{\pi}{2}\frac{Q_{\rm c}}{Q}\right)^2 ,
& Q/Q_{\rm c} \rightarrow \infty
\end{array} \right .
\end{eqnarray}
The SL state energy per area is represented by the dots in
Fig.~\ref{hannafig1}.

Although it is not needed for calculating the stiffnesses or
susceptibility of the SL, the SL solution for
$\tilde{\theta}({\bf r})$ is given by\cite{fetter}
\begin{equation}
\label{eq:thetasoln}
\sin[\frac{1}{2}(\tilde{\theta}-\pi)] =
{\rm sn}({\bf\hat{e}}_1\cdot ({\bf r} - {\bf r}_0)/\eta\xi,\eta) ,
\end{equation}
where sn denotes the sine-amplitude Jacobian elliptic function
with parameter $\eta$.\cite{gr}
Near the CI transition, when $Q\rightarrow Q_{\rm c}$,
$\tilde{\theta}$ is very close to being a periodic superposition of
single-soliton solutions, spaced apart by $\bar{\bf L}_{\rm s}$,
so that $\tilde{\theta}({\bf r}) \approx
\sum_j \tilde{\theta}_{\rm ss}({\bf r} - j\bar{\bf L}_{\rm s})$,
where $\tilde{\theta}_{\rm ss}({\bf r})$ denotes the single-soliton
solution, Eq.~(\ref{eq:sol}).
In the SL state, especially away from the CI transition,
it is sometimes useful to work with
\begin{equation}
\label{eq:thetas}
\theta_{\rm s}({\bf r}) \equiv
\tilde{\theta}({\bf r}) - \bar{\bf Q}_{\rm s} \cdot {\bf r} =
\theta({\bf r}) + ({\bf Q} - \bar{\bf Q}_{\rm s}) \cdot {\bf r} ,
\end{equation}
because it is periodic in the SL spacing, so that
$\theta_{\rm s}({\bf r} + \bar{\bf L}_{\rm s}) = \theta_{\rm s}({\bf r})$,
where
$\bar{\bf L}_{\rm s} \equiv 
{\bf\hat{Q}}_{\rm s}2\pi/\bar{Q}_{\rm s}$.
In the limit $Q/Q_{\rm c}\rightarrow\infty$,
$\bar{\bf Q}_{\rm s} \rightarrow {\bf Q}$ and
$\theta_{\rm s} \rightarrow \theta \rightarrow 0$,
so that we may regard $\theta_{\rm s}({\bf r})$ as a small quantity.
Expressing the sine-Gordon equation [Eq.~(\ref{eq:sg2})] in terms of
$\theta_{\rm s}$ and working to lowest order in $\theta_{\rm s}$ gives
${\bf \nabla}^2 \theta_{\rm s} \approx
(1/\xi^2) \sin [\bar{\bf Q}_{\rm s} \cdot
({\bf r} - {\bf r}_0)]$,
so that
\begin{equation}
\label{eq:thetabigq}
\lim_{Q\rightarrow\infty}
\theta_{\rm s}({\bf r}) \approx
-\left(\frac{\pi}{4} \frac{Q_{\rm c}}{\bar{Q}_{\rm s}} \right)^2
\sin [\bar{\bf Q}_{\rm s} \cdot ({\bf r} - {\bf r}_0)] .
\end{equation}

\section{Stiffnesses of the Soliton Lattice}
\label{sec:stiff}
 
The elastic constants of the soliton lattice are given by
the stiffness tensor $K_{ij}$.
The change in the energy due to varying the spacing
between the soliton lines is characterized by
the compressional stiffness $K_{11}$.
The shear stiffness $K_{22}$ is associated with the change in energy due to
shearing the upper and lower ends of the soliton lines
in opposite directions,
and is equivalent to a combined rotation and compression.
We use the calculated stiffnesses to describe the $B_\parallel$-dependence
of the KT temperature\cite{read} for the dislocation-mediated KT melting
transition\cite{dennijs}
of the soliton-lattice.

We calculate the stiffness tensor by two methods.
First, we calculate the stiffness $K_{ij}$ that is obtained
by differentiating ${\cal E}/L_1L_2$ in Eq.~(\ref{eq:elxly})
with respect to the $i,j$ components of ${\bf Q}_{\rm s}$ for
fixed ${\bf Q}$ at the extremal, where Eq.~(\ref{eq:eom})
is zero.
Then we calculate the stiffness tensor $\tilde{K}_{ij}$ obtained
by calculating the effects of fluctuations of the angle variable
$\tilde{\theta}({\bf r})$ away from its ground-state value,
Eq.~(\ref{eq:thetasoln}).

We begin by calculating the stiffness tensor $K_{ij}$ from the dependence of
the ground-state energy [Eq.~(\ref{eq:elxly})] on the soliton-lattice
wave vector $Q_{\rm s}$.
The expressions we obtain for $K_{ij}$ by this method
have been carried out for all values of $Q \ge Q_{\rm c}$,
and agree with the results obtained in Ref.~\onlinecite{read},
in the limit $Q \rightarrow Q_{\rm c}$.
Because the stiffnesses involve the second derivative of
${\cal E}/L_1L_2$ with respect to
the components of ${\bf Q}_{\rm s}$ at fixed ${\bf Q}$,
the terms in Eq.~(\ref{eq:elxly}) that depend explicitly on ${\bf Q}$
(including the term $-\rho_s{\bf Q}\cdot{\bf Q}_s$ that gives the
orientational dependence of the energy per area)
do not contribute to $K_{ij}$.
Thus
\begin{eqnarray}
\label{eq:kij}
K_{ij} &=&
\lim_{{\bf Q}_{\rm s}\rightarrow \bar{\bf Q}_{\rm s}}
\frac{\partial^2}{\partial Q_{{\rm s}i} \partial Q_{{\rm s}j}}
\left ( \frac{{\cal E}}{L_1L_2} \right )_{\bf Q}
\\ \nonumber &=&
\rho_{\rm s} Q_{\rm c} \lim_{{\bf Q}_{\rm s}\rightarrow \bar{\bf Q}_{\rm s}}
\frac{\partial}{\partial Q_{{\rm s}i}}
\left [ \frac{Q_{{\rm s}j}}{Q_{\rm s}} \frac{E(\eta)}{\eta} \right ]_{\bf Q}
\nonumber \\ &=&
\rho_{\rm s} \lim_{{\bf Q}_{\rm s}\rightarrow \bar{\bf Q}_{\rm s}}
\left [
\left ( \delta_{ij} -
\frac{\bar{Q}_{{\rm s}i}\bar{Q}_{{\rm s}j}}
     {\bar{Q}_{\rm s}^2} \right )
\frac{1}{\bar{Q}_{\rm s}} +
\frac{\bar{Q}_{{\rm s}i}\bar{Q}_{{\rm s}j}}{\bar{Q}_{\rm s}^2}
\frac{\partial}{\partial Q_{\rm s}}
\right ]
\\ \nonumber
&& \qquad \qquad \qquad \times
\left [ Q_{\rm c} \frac{E(\eta)}{\eta} \right ]_{\bf Q} ,
\end{eqnarray}
where we have used Eqs. (\ref{eq:eom}) and (\ref{eq:qoqc}).
Since $\bar{Q}_{{\rm s}2}=0$, it follows that $K_{12} = K_{21} = 0$.

Using the results of Sec.~\ref{sec:solat},
and the identity\cite{gr}
\begin{equation}
\label{eq:dkdn}
\frac{dK}{d\eta} = \frac{E(\eta)}{\eta(1-\eta^2)} -
                   \frac{K(\eta)}{\eta} ,
\end{equation}
from which it follows that
\begin{equation}
\label{eq:detadqs}
\left(\frac{d\eta}{dQ_{\rm s}}\right)_Q =
-\frac{(\pi/2)^2}{Q_{\rm c}} \frac{(1-\eta^2)}{E(\eta)}
\left( \frac{Q_{\rm c}}{Q_{\rm s}} \right)^2 ,
\end{equation}
we find that the compressional elastic constant
$K_{11}$ is equal to
\begin{eqnarray}
\label{eq:k11}
&& \frac{K_{11}}{\rho_{\rm s}} = 
\lim_{{\bf Q}_{\rm s}\rightarrow \bar{\bf Q}_{\rm s}}
\frac{\partial}{\partial\bar{Q}_{\rm s}}
\left [ Q_{\rm c} \frac{E(\eta)}{\eta} \right ]_{\bf Q}
= \frac{\partial Q}{\partial\bar{Q}_{\rm s}}
\\ \nonumber &&
= \frac{16(1-\bar{\eta}^2)}{(\bar{Q}_{\rm s}\xi)^3 (Q\xi) \bar{\eta}^4}
\rightarrow \left \lbrace
\begin{array}{cc}
(2/\pi^2) \epsilon \ln^2 (1/\epsilon) ,
& Q/Q_{\rm c} \rightarrow 1 \\
1 - \frac{3}{32} \left ( \frac{\pi}{4} \frac{Q_{\rm c}}{Q} \right )^4 ,
& Q/Q_{\rm c} \rightarrow \infty
\end{array} \right .
\end{eqnarray}
In the limit $Q/Q_{\rm c} \rightarrow 1$ when the soliton lines
are far apart, $K_{11}$ is very small [of order
$\epsilon\sim\exp(-L_{\rm s}/\xi)$, see Eq.~(\ref{eq:qsofq})].
The energy cost of compression very close to the CI transition
is due to the exponentially weak intersoliton interaction energy.
The energy per area due to the string tension of the soliton lines
(the term $\rho_s Q_c Q_s$) does not contribute to $K_{11}$,
although it does contribute to $K_{22}$, as we explain below.

As explained in Ref.~\onlinecite{copper},
soliton lines meander at finite temperature
and are no longer straight;
collisions between meandering soliton lines produce an
effective entropic repulsion between the solitons that
dominates the exponential repulsion
at any nonzero temperature, for $L_{\rm s}/\xi$ sufficiently large.
This effect renormalizes the compressional stiffness $K_{11}$
upwards so that it becomes proportional to $T^2$.
In the limit $Q/Q_{\rm c} \rightarrow \infty$,
the tunneling term in the PT energy (\ref{eq:tp2}) fluctuates
on very short length scale and averages to zero, so that
Eq.~(\ref{eq:tp2}) becomes the isotropic XY model;
thus one expects $K_{11}$ to approach the
pseudospin stiffness $\rho_{\rm s}$ in that limit.

{}From Eqs. (\ref{eq:qoqc}) and (\ref{eq:kij}), it follows that
the shear elastic constant $K_{22}$ is given by
\begin{equation}
\label{eq:k22}
\frac{K_{22}}{\rho_{\rm s}} = \frac{Q}{\bar{Q}_{\rm s}}
\rightarrow \left \lbrace
\begin{array}{cc}
(2/\pi^2) \ln (1/\epsilon) ,
& Q/Q_{\rm c} \rightarrow 1 \\
1 + \frac{1}{2} \left ( \frac{\pi}{4} \frac{Q_{\rm c}}{Q} \right )^4 ,
& Q/Q_{\rm c} \rightarrow \infty
\end{array} \right .
\end{equation}
As expected, the shear stiffness $K_{22}$
approaches the pseudospin stiffness $\rho_{\rm s}$
in the limit $Q/Q_{\rm c} \rightarrow \infty$.
But in the limit $Q\rightarrow Q_{\rm c}$,
$K_{22}$ diverges as $Q_{\rm c}/\bar{Q}_{\rm s}$.
The origin of this effect is that the shear motion
described by $K_{22}$ increases the total length of the soliton lines.
By definition, $K_{\rm 22}$ describes a shear displacement
in which $Q_{{\rm s}2}$ changes, while $Q_{{\rm s}1}$ remains fixed:
i.e., the solitons lines tilt away
from their equilibrium ``vertical'' (${\bf \hat{e}}_2$) direction
by a small angle \mbox{$\phi=Q_{{\rm s}2}/Q_{\rm s}$},
while keeping their ``horizontal separation''
${\bf \hat{e}}_1\cdot{\bf L}_{\rm s}=\bar{L}_{\rm s}$ constant.
This shear motion reduces the mean soliton separation
(the separation along the direction perpendicular to the tilted soliton lines)
to \mbox{$L_{\rm s}=\bar{L}_{\rm s}\cos\phi$}, so that
\mbox{$Q_{\rm s}\rightarrow{\bar Q}_{\rm s}/\cos\phi$}.
The $K_{22}$-shear corresponds to a global rotation
plus a compression of the SL.
Packing the solitons lines more closely together in a fixed
sample area produces more total soliton line length,
which costs more soliton-line creation energy.

Because the term $-\rho_s {\bf Q} \cdot {\bf Q}_s$ that contains the
orientational dependence of the energy per area
[Eq.~(\ref{eq:elxly})] is linearly proportional to $Q_{s1}$,
it cannot contribute to the stiffnesses (which are proportional to
second derivatives of the energy per area with respect to the
components of ${\bf Q}_s$) at all.
It might be supposed that the rotation of the soliton lines that occurs
upon shearing should cost energy, but this is not so, because
a shear is a combination of rotation and compression, rather than
a pure rotation.
The creation energy per area of the SL near the CI transition
($Q\approx Q_c$) is
$\rho_{\rm s} (Q_{\rm c} Q_{\rm s}-{\bf Q}\cdot{\bf Q}_{\rm s})$,
and consists of two terms.
The first term ($\rho_{\rm s} Q_{\rm c} Q_{\rm s}$)
is just the total line length of the solitons times the line tension,
divided by the total area.
The second term ($-\rho_{\rm s}{\bf Q}\cdot{\bf Q}_{\rm s}$)
depends explictly on the angle $\phi$ between ${\bf Q}$ and ${\bf Q}_{\rm s}$,
and determines the orientation of the SL because it is minimized
by choosing ${\bf Q}_{\rm s}$ along $Q$ (i.e., $\phi=0$); thus,
a different choice of SL orientation (i.e., a pure rotation of the soliton
lines) would cost more energy.
Interestingly, the second term in the creation energy is unchanged by a
shear, because the energy cost of rotating the soliton lines is exactly offset
by the increase in total soliton line length:
$-[Q({\bar Q}_{\rm s}/\cos\phi)]\cos\phi = -Q \bar{Q}_{\rm s}$,
independent of $\phi$.
The only change in the creation energy comes from the first term,
which depends only on the density of soliton lines:
\mbox{$\rho_{\rm s} Q_{\rm c} (\bar{Q}_{\rm s}/\cos\phi)$}.
Sufficiently close to the CI transition (i.e., when $L_{\rm s}/\xi\gg 1$),
the exponentially small interactions may be neglected in comparison
to the creation energy.
For $Q\rightarrow Q_{\rm c}$
and small shear ($Q_{{\rm s}2}/Q_{{\rm s}1} \ll 1$),
\begin{eqnarray}
\label{eq:eci}
\frac{{\cal E}}{L_1L_2} &\rightarrow&
\rho_{\rm s} Q_{\rm c} Q_{\rm s} - t/2\pi\ell^2
\\ \nonumber
&\approx& \rho_{\rm s} Q_{\rm c} \left[
Q_{\rm s} + \frac{1}{2} \frac{Q_{{\rm s}2}^2}{Q_{\rm s}} \right ]
- t/2\pi\ell^2 ,
\end{eqnarray}
so that
$K_{22}\rightarrow\rho_{\rm s}
Q_{\rm c}/{\bar Q}_{\rm s}\rightarrow\infty$
as $Q\rightarrow Q_{\rm c}$, in agreement with the results of
Ref.~\onlinecite{read}.

\begin{figure}[t]
\epsfxsize4.0in
\centerline{\epsffile{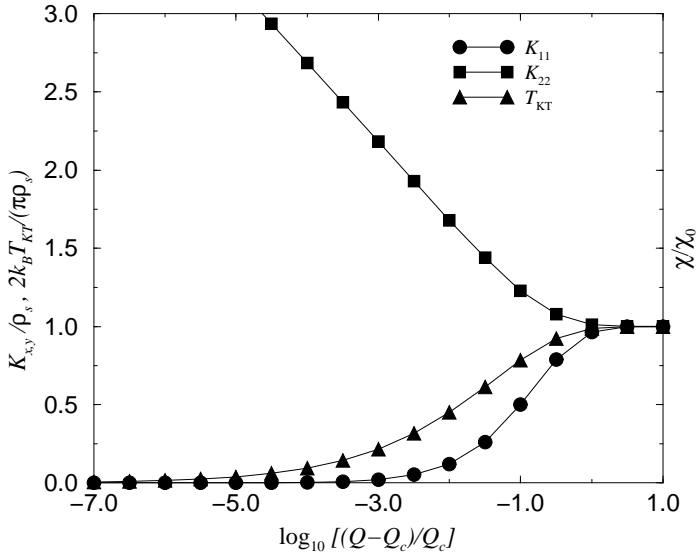}}
\caption{Soliton-lattice stiffnesses $\rho_x \equiv K_{11}$
and $\rho_y \equiv K_{22}$, and their geometric mean, which
is proportional to the Kosterlitz-Thouless temperature, $T_{\rm KT}$.}
\label{hannafig5}
\end{figure}

The fact that bilayer phase-coherent 2LQH states can exhibit
a finite-temperature Kosterlitz-Thouless (KT) transition
in the absence of interlayer tunneling ($t\rightarrow 0$) has
been discussed in earlier work.\cite{yangprl,moonbigprb}
Although finite $t$ removes the possibility of a KT transition
in the commensurate phase of 2LQH systems
by altering the nature of the long-range interaction between
vortices (``merons'' in this case),\cite{yangbigprb}
the SL phase does support a finite-temperature KT transition
due to dislocation-mediated melting of the SL.\cite{dennijs}
As discussed in Ref.~\onlinecite{read},
the KT temperature may be estimated as
\begin{eqnarray}
\label{eq:tkt}
&& \frac{k_B T_{\rm KT}}{(\pi/2)\rho_{\rm s}} \sim
\frac{1}{\rho_{\rm s}} \sqrt{\det (K_{ij})} =
\frac{1}{\rho_{\rm s}} \sqrt{K_{11}K_{22}}
\\ \nonumber && \quad \sim
\frac{4\sqrt{1-\bar{\eta}^2}}{(\bar{Q}_{\rm s}\xi)^2 \bar{\eta}^2}
\rightarrow \left \lbrace
\begin{array}{cc}
(2/\pi^2) \sqrt{\epsilon \ln^3 (1/\epsilon)} ,
& Q/Q_{\rm c} \rightarrow 1 \\
1 + \frac{13}{64} \left ( \frac{\pi}{4} \frac{Q_{\rm c}}{Q} \right )^4 ,
& Q/Q_{\rm c} \rightarrow \infty
\end{array}
\right . 
\end{eqnarray}
where we have used the zero-temperature values for $K_{ij}$ that we
calculated previously to make a rough estimate the KT temperature.
As mentioned earlier, finite-temperature fluctuation effects
can strongly renormalize $K_{11}$,\cite{copper}
and may also effect $K_{22}$.
Our results agree with those of Ref.~\onlinecite{read}
in the limit $Q \rightarrow Q_{\rm c}$.
We have plotted the compressional ($K_{11}$) and
shear ($K_{22}$) stiffnesses in Fig.~\ref{hannafig5},
together with the KT temperature.

The KT transition would be most easily measured from the temperature
dependence of the linear response to oppositely directed currents
in each layer.
This would require double-layer electron devices with layers that
could be contacted separately.
Unfortunately, the leakage currents produced when the interlayer tunneling
is not vanishingly small would make it difficult, perhaps impossible,
to set up oppositely directed currents in each layer.
However, because the SL dislocations are electrically charged,
it might be possible that the KT transition could be signalled
by an increase in the usual longitudinal resistivity $\rho_{xx}(T)$,
measured in devices with the current flowing in the
same direction in both layers.
The increase in $\rho_{xx}(T)$ would originate from the proliferation
of unbound charged dislocations above the transition temperature.

We now calculate an alternate stiffness tensor $\tilde{K}_{ij}$
by examining the effect of deviations of the angle variable
$\tilde{\theta}({\bf r})$ from its ground-state value.
We write
$\tilde{\theta}({\bf r})=
\tilde{\theta}_0({\bf r})+\delta\tilde{\theta}({\bf r})$,
where $\tilde{\theta}_0({\bf r})$ is the ground-state solution
that minimizes the PT energy (\ref{eq:tp2})
and therefore satisfies Eq.~(\ref{eq:sg1}),
and $\delta\tilde{\theta}({\bf r})$ is the deviation of $\tilde{\theta}$
from its ground-state value.
We do not include dynamics here, because our focus is on
ground-state, rather than excited-state, properties.
The PT energy for $\tilde{\theta}$ is
$E_{\rm PT}[\tilde{\theta}_0+\delta\tilde{\theta}]
\equiv E_{\rm PT}[\tilde{\theta}_0] + \delta H$,
where $E_{\rm PT}$, given by Eq.~(\ref{eq:tp2}), is the PT energy from
which the ground-state $\tilde{\theta}_0({\bf r})$ is determined
via Eq.~(\ref{eq:sg1}), and
\begin{eqnarray}
\label{eq:delh}
\delta H &\approx&
\frac{1}{2} \int \frac{d^2r}{2\pi\ell^2}
\left[ t\cos\tilde{\theta}_0 \delta\tilde{\theta}^2 +
2\pi\ell^2 \rho_{\rm s}
\left({\bf \nabla}\delta\tilde{\theta}\right)^2
\right]
\\ \nonumber
&=&
\frac{1}{2} \int \frac{d^2r}{2\pi\ell^2} \delta\tilde{\theta}
\left[ t\cos\tilde{\theta}_0  -
2\pi\ell^2 \rho_{\rm s}
{\bf \nabla}^2
\right] \delta\tilde{\theta} ,
\end{eqnarray}
where we have kept terms up to quadratic order in $\delta\tilde{\theta}$.
There are no terms linear in $\delta\tilde{\theta}$
because $\tilde{\theta}_0({\bf r})$ is determined by minimizing
$E_{\rm PT}$ with respect to variations in $\tilde{\theta}$.
The total energy
is minimized by choosing $\delta\tilde{\theta}$
from among the  eigenvalues of the bracketed Schr\"odinger-like
operator in Eq.~(\ref{eq:delh}), so that
\begin{equation}
\label{eq:scheq}
\left[ t\cos\tilde{\theta}_0  -
2\pi\ell^2 \rho_{\rm s}
{\bf \nabla}^2
\right] \delta\tilde{\theta}_{\bf q}
= {\cal E}_{\bf q} \delta\tilde{\theta}_{\bf q}
\end{equation}

If we take ${\bf B}_\parallel = B_\parallel{\bf \hat{y}}$
so that ${\bf Q} = Q {\bf \hat{x}}$,
then $\tilde{\theta}_0({\bf r})$ depends only on $x$,
$\delta H$ is translationally invariant in the $y$ direction,
and we may write
$\delta\tilde{\theta}_{\bf q}({\bf r})
= \exp(iq_yy) \delta\tilde{\theta}_{q_x}(x)$.
The term $t\cos\tilde{\theta}_0$ in Eq.~(\ref{eq:scheq})
is periodic in the $x$ direction, with a period of $\bar{L}_s$.
As shown in Ref.~\onlinecite{lebwohl},
when $\tilde{\theta}_0({\bf r})$ has the form (\ref{eq:thetasoln}),
Eq.~(\ref{eq:scheq}) becomes  Lam\'e's equation, after a simple
rescaling of $x$.
Lam\'e's equation has three simple solutions, two of which have
low-energy limits.
The first type of solution has zero energy and
corresponds to a uniform translation of the vortex lines,
$\delta\tilde{\theta} \propto \partial\tilde{\theta}_0/\partial x_0$,
where $x_0$ is the $x$ component of ${\bf r}_0$ in Eq.~(\ref{eq:thetasoln}).
Of greatest interest to us are the
type of solutions which have been called ``vortex oscillations''
in the context of long Josephson junctions in parallel magnetic
fields.\cite{fetter}
{}From Ref.~\onlinecite{fetter}, it follows that in the long-wavelength
limit,
\begin{equation}
\label{eq:ktilde}
\frac{\delta {\cal E}}{L_1L_2} = \frac{1}{2} \left(
K_{11} q_x^2 + \rho_{\rm s} q_y^2 \right) ,
\end{equation}
where $K_{11}$ is equal to the compressional stiffness in
Eq.~(\ref{eq:k11}), and $q_x$ is the crystal momentum along the
$x$ direction.

Although $\tilde{K}_{11}=K_{11}$, we find that
$\tilde{K}_{22}=\rho_{\rm s}\ne K_{22}$.
The reason for the discrepancy between $\tilde{K}_{22}$ and
$K_{22}$ is not obvious.
It may be that using $K_{22}=Q/Q_{\rm s}$ for the transverse stiffness
is valid only at very long wavelengths $q_y < 1/Y_{SL}$,\cite{readpriv}
where $Y_{SL}=L_{\rm s}\sqrt{K_{22}/K_{11}}$ is Read's estimate
of the minimum $y$-distance for SL dislocations to interact
logarithmically.\cite{read}
It is also possible that $\tilde{K}_{22}$, which is based on a calculation
that assumes $\delta\tilde{\theta}$ is everywhere small,
is not able to describe uniform shear, which can move solitons lines far
from their equilbrium positions on the scale of the soliton line thickness
$\xi$.
The relationship between $\tilde{K}_{22}$ and $K_{22}$
requires further clarification, including a stronger argument
for preferring $K_{22}$ over $\tilde{K}_{22}$ in estimating
the KT temperature.

\section{In-Plane Magnetization}
\label{sec:mag}

The interlayer phase coherent 2LQH state exhibits an in-plane
magnetization ${\bf M}_\parallel$ in the presence of an in-plane
magnetic field ${\bf B}_\parallel$.
The in-plane magnetization can be calculating by differentiating
the minimized ground-state energy per volume with respect to the
parallel magnetic field:
\begin{equation}
\label{eq:magpar}
{\bf M}_\parallel =
-\frac{1}{L_1L_2d} \frac{\partial\bar{{\cal E}}}
                       {\partial {\bf B}_\parallel} =
-\frac{2\pi}{\phi_0} {\bf \hat{z}\times}
\frac{\partial}{\partial {\bf Q}}
\left ( \frac{\bar{{\cal E}}}{L_1L_2} \right ) ,
\end{equation}
where $\bar{{\cal E}}/L_1L_2$ is given by Eq.~(\ref{eq:emin}).
In order to carry out the differentiation in Eq.~(\ref{eq:magpar}),
we first differentiate Eq.~(\ref{eq:qoqc}) with respect to $Q$
and make use of Eq.~(\ref{eq:qsoqc}) to obtain
\begin{equation}
\label{eq:detadq}
\frac{\partial\bar{\eta}}{\partial Q} =
-\left(\frac{2}{\pi}\right)^2 \bar{\eta}^3
\frac{\bar{Q}_{\rm s}}{Q_{\rm c}^2}
\left(1+Q^2\ell^2/4\right) .
\end{equation}
We note here that the $Q^2\ell^2/4$ term in Eq.~(\ref{eq:detadq})
arises from differentiating $Q_{\rm c}$ in Eq.~(\ref{eq:qoqc})
with respect to $Q$;
the $Q$-dependence of $Q_{\rm c}$ is due to the dependence of
the tunneling matrix element on the parallel magnetic field,
$t=t_0\exp (-Q^2\ell^2/4)$, which is a single-particle effect.\cite{hu}
The tunneling part of the equilibrium energy per area
will also give a contribution to the in-plane magnetization
proportional to $\partial t/\partial Q$, again due to dependence
of $t$ on $Q$.
It is convenient to separate ${\bf M}_\parallel$ into two parts:
\mbox{${\bf M}_\parallel \equiv {\bf M}_{\rm SL} + {\bf M}_{\rm t}$},
where ${\bf M}_{\rm SL}$ is calculated at fixed $t$
($t$ independent of $Q$),
and ${\bf M}_{\rm t}$ arises from the $Q$-dependence of $t$.

Using Eqs. (\ref{eq:magpar}) and (\ref{eq:detadq}) it is
straightforward to show that the SL contribution to
the parallel magnetization is
\begin{equation}
\label{eq:magsl}
{\bf M}_{\rm SL} =
-M_0 {\bf \hat{z}\times} ({\bf Q} - \bar{{\bf Q}}_{\rm s})/Q_{\rm c} ,
\end{equation}
where 
\begin{equation}
M_0\equiv 2\pi\rho_{\rm s} Q_{\rm c}/\phi_0
\sim 0.5 {\rm A/m}
\end{equation}
sets the scale of the SL magnetization,
and the numerical estimate of $M_0$ is given for the ``typical'' GaAs
sample described in Sec.~\ref{sec:singsols}.
For such a sample,
\begin{equation}
M_0d = 2\pi\rho_{\rm s} Q_{\rm c}d/\phi_0 \sim 10^{-8}~{\rm A}
\end{equation}
sets the scale of the SL magnetic moment per unit area.
Thus magnetometers with sensitivities in the range of
$10^{-14}$~Am$^2$ to $10^{-12}$~Am$^2$ would
require sample areas in the range of 1~mm$^2$ and 1~cm$^2$
to measure $M_{\rm SL}$.
The magnitude of the SL magnetization behaves like
\begin{equation}
\label{eq:mslbehav}
\left|\frac{M_{\rm SL}}{M_0}\right| \rightarrow \left\lbrace
\begin{array}{cc}
1 - (\pi^2/2)/\ln(1/\epsilon) ,
& Q/Q_{\rm c}\rightarrow 1 \\
\frac{\pi}{64} \left(\frac{\pi}{2}\frac{Q_{\rm c}}{Q}\right)^3 ,
& Q/Q_{\rm c}\rightarrow\infty
\end{array} \right .
\end{equation}

The SL magnetization may also be calculated directly from the
pseusdospin supercurrent density,\cite{moonbigprb}
${\bf J}_{zz} \equiv {\bf J}_1 - {\bf J}_2
= (2\rho_{\rm s}/\hbar)\nabla\theta$,
and the definition of the magnetic moment.
The electrical current ${\bf I}$ in layers 1 and 2 is
\begin{equation}
\label{eq:e1def}
{\bf I} = {\bf I}_1 = -{\bf I}_2 = L_y (e\rho_{\rm s}/\hbar)
                 (\nabla \tilde{\theta} - {\bf Q}) .
\end{equation}
The magnetization produced by the above current is
therefore\cite{jackson}
\begin{eqnarray}
\label{eq:magslalt}
{\bf M}_{\rm SL} &=&
\frac{1}{L_xL_yd} {\bf \hat{z}\times}  \int\!\!\int \frac{\bf I}{c} dx dz
\\ \nonumber
&=& \frac{e\rho_{\rm s}}{\hbar} {\bf \hat{z}\times} \frac{1}{L_x}
  \int_0^{L_x} (\nabla \tilde{\theta} - {\bf Q}) dx
\\ \nonumber
&=& -\frac{e\rho_{\rm s}}{\hbar}{\bf \hat{z}\times}
                          ({\bf Q} - \bar{\bf Q}_{\rm s}) ,
\end{eqnarray}
in agreement with Eq.~(\ref{eq:magsl}).

The 2LQH interlayer phase coherent state may be regarded as
a pseudospin-channel superconductor.\cite{moonbigprb}
The magnetization ${\bf M}_\parallel$ is due to
pseudospin supercurrents, corresponding to electrical
currents traveling in opposite directions in each layer,
which partially screen ${\bf B}_\parallel$.
For $Q<Q_c$, ${\bf B}_\parallel$ is maximally excluded
from the region between the planes;
but when $Q\ge Q_c$, additional $B_\parallel$ penetrates the region
between the plates in the form of solitons that each carry a single
flux quantum $\phi_0=h/e$ (corresponding to a phase change
$\Delta\tilde{\theta}=2\pi$), leading to a precipitous decline in the
magnetization.
The direction of ${\bf M}_\parallel$ is opposite to ${\bf B}_\parallel$,
in accord with Lenz's law.
An exactly analogous effect occurs for magnetic fields
applied parallel to narrow insulating regions (Josephson junctions)
between superconductors.

The contribution to the magnetization due to
the $Q$-dependence of the tunneling matrix element $t$ is
\begin{eqnarray}
\label{eq:magtun}
{\bf M}_{\rm t} &=&
-\frac{1}{2} \left(\frac{\ell}{\xi}\right)^2
M_0 {\bf \hat{z}\times} \left[
\left(\frac{2}{\bar{\eta}^2}-1\right) \frac{\bf Q}{Q_{\rm c}}
\right.
\\ \nonumber && \left. \qquad \qquad \qquad \qquad
- \frac{1}{2} \left(\frac{4}{\pi} \frac{Q}{Q_c}\right)^2
\frac{\bar{\bf Q}_{\rm s}}{Q_{\rm c}}
\right] .
\end{eqnarray}
In the commensurate phase,
${\bf M}_{\rm t} = -(M_0/2) (\ell/\xi)^2 {\bf \hat{z}\times}
{\bf Q}/Q_{\rm c}$,
and the magnitude of the tunneling contribution to the parallel
magnetization behaves like
\begin{equation}
\label{eq:mtbehav}
\left|\frac{M_{\rm t}}{M_0}\right| \rightarrow
\frac{1}{2} \left(\frac{\ell}{\xi}\right)^2
\left\lbrace
\begin{array}{cc}
1 - 4/\ln(1/\epsilon) ,
& Q/Q_{\rm c}\rightarrow 1 \\
\frac{1}{2} (\pi/4)^2 (Q_{\rm c}/Q) ,
& Q/Q_{\rm c}\rightarrow\infty
\end{array} \right .
\end{equation}

\begin{figure}[b]
\epsfxsize4.0in
\centerline{\epsffile{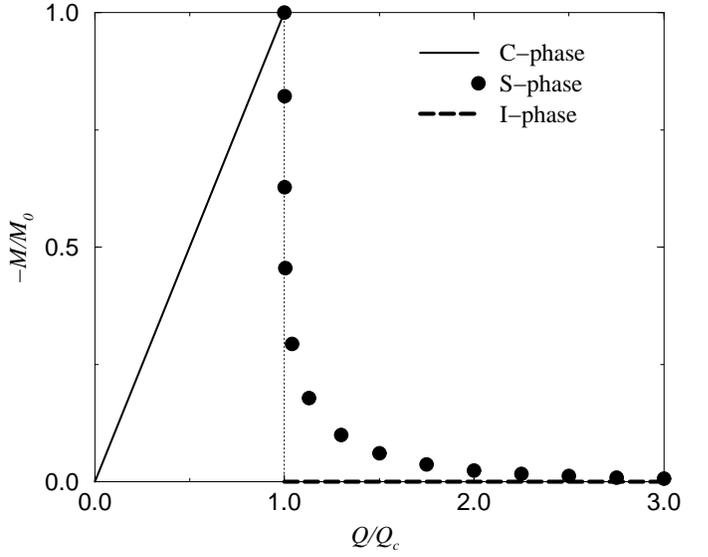}}
\caption{SL contribution to the magnetization,
which is proportional to $(Q-\bar{Q}_{\rm s})$
and therefore drops precipitously for $Q > Q_{\rm c}$.}
\label{hannafig6}
\end{figure}

The SL magnetization is plotted in Fig.~\ref{hannafig6}.
It is useful to compare the total SL magnetization
$M_0$ to the scale of the Landau diamagnetism in a $\nu_T=1$
2LQH system:
\begin{equation}
\label{eq:m0}
\frac{M_0}{(n/d) \mu_{\rm B}^*} =
\frac{M_0\phi_0 d}{\mu_{\rm B}^* B_\perp} =
16 \frac{d}{\xi} \frac{\rho_{\rm s}}{\hbar\omega_{\rm c}}
\sim 0.1 ,
\end{equation}
where $\mu_{\rm B}^* = e\hbar/2m^*$ is the effective Bohr magneton,
and we have made use of the parameters for the ``typical'' GaAs
sample described in Sec.~\ref{sec:singsols}.
This shows that the SL magnetization is expected to
be roughly an order of magnitude smaller than the Landau diamagnetism.
It is interesting to note that the weak signals associated with
orbital diamagnetism in two-dimensional electron systems
have been measured at high magnetic fields
by torsional magnetometry,\cite{tormag}
SQUID magnetometry,\cite{squid}
and micromechanical cantilever magnetometry.\cite{cantilev}
The torque on the 2LQH sample
in the presence of both a perpendicular and parallel magnetic field
has both a Landau-diamagnetic component
$\tau_\perp = (L_xL_yd) M_\perp B_\parallel$ and a SL component,
$\tau_\parallel = (L_xL_yd) M_\parallel B_\perp$.
The smallness of the parallel moment $M_\parallel$ is offset by the
large perpendicular magnetic field $B_\perp$ in the expression for
the torque $\tau_\parallel$.
The SL magnetization ${\bf M}_{\rm SL}$ might also be measured using
high-field magnetometry techniques mentioned
above.

It follows from Eq.~(\ref{eq:magpar}) that when $M_\parallel$ is plotted
against $B_\parallel$, the area under under the resulting curve
from $B_\parallel=0$ to $B_\parallel=\infty$ is
\begin{eqnarray}
\int_0^\infty M_\parallel dB_\parallel
&=& \frac{1}{L_xL_yd} \left[
\bar{\cal E}(B_\parallel=\infty) - \bar{\cal E}(B_\parallel=0) \right]
\\ \nonumber
&=& \left(\frac{t \langle m_x \rangle}{2\pi\ell^2d}\right)_{B_\parallel=0} , 
\end{eqnarray}
where $\langle m_x \rangle$ is the ground-state expectation value of
the $x$ component of the pseudospin order parameter, which has a
value of $\sqrt{1-m_z^2}$ in the HFA, so that $\langle m_x \rangle=1$
(in the HFA) when the layers are balanced ($m_z=0$).
Thus the area under the $M_\parallel$ versus $B_\parallel$ curve may
be regarded either as a measurement of $t$ (if quantum fluctuations
in the ground state are neglected), or as a measure of order-parameter
suppression (of $m_x$) due to quantum fluctuations\cite{schliemann}
(if $t$ can be measured separately).

\begin{figure}[t]
\epsfxsize4.0in
\centerline{\epsffile{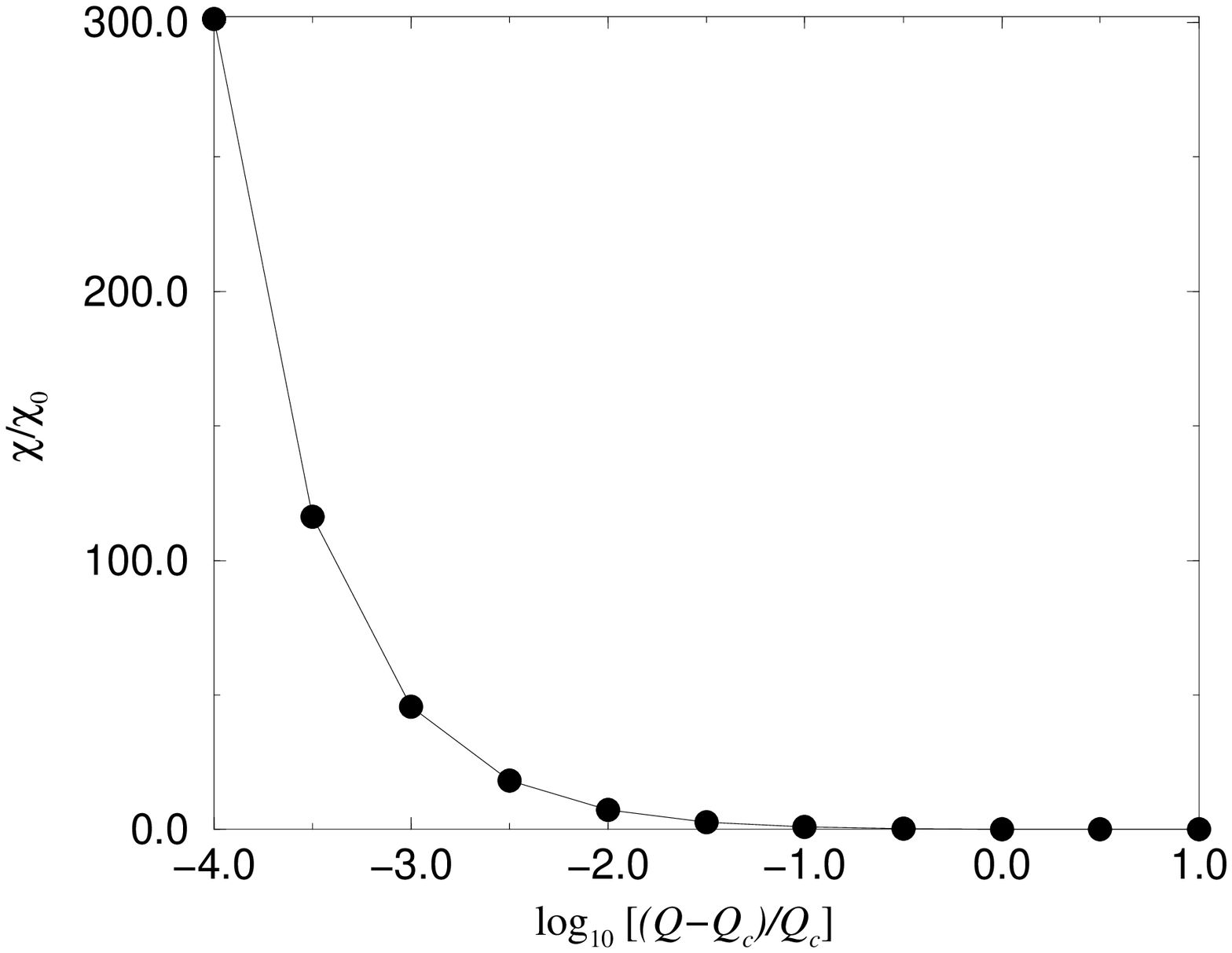}}
\caption{SL contribution to the magnetic susceptibility,
which diverges as $1/(Q-Q_{\rm c})\ln[Q_{\rm c}/(Q-Q_{\rm c})]$
for $Q\rightarrow Q_{\rm c}$.}
\label{hannafig7}
\end{figure}

Equations (\ref{eq:magsl}) and (\ref{eq:qsofq}) show that
${\bf M}_{\rm SL} = -M_0 {\bf \hat{z}\times}{\bf Q}/Q_c =
-(\chi_0/\mu_0) {\bf B}_\parallel$
in the commensurate phase ($Q< Q_{\rm c}$),
where $\chi_0 \equiv \mu_0 (2\pi/\phi_0)^2 \rho_{\rm s} d$
sets the scale of the SL contribution to the magnetic susceptibility.
The SL contribution to the parallel-field magnetic susceptibility
is defined for fixed $t$ (independent of $Q$):
\begin{eqnarray}
\label{eq:chi}
\chi_{\rm SL} & \equiv & \mu_0
\left(\frac{\partial M_{\rm SL}}{\partial B_\parallel}\right)_t = \mu_0
\frac{2\pi d}{\phi_0} \left(\frac{\partial M_{\rm SL}}
                                {\partial Q}\right)_t
\\ \nonumber &=&
\chi_0 \left ( 1 -
\frac{\partial \bar{Q}_{\rm s}}{\partial Q}
\right )_t
= \chi_0 \left ( 1 - \frac{\rho_{\rm s}}{K_{11}} \right )
\\ \nonumber
&\rightarrow& \chi_0 \left \lbrace
\begin{array}{cc}
1 ,
& Q/Q_{\rm c} < 1 \\
-(\pi^2/2) / [\epsilon \ln (1/\epsilon)] ,
& Q/Q_{\rm c} \rightarrow 1^+ \\
-\frac{3}{2} \left ( \frac{\pi}{4} \frac{Q_{\rm c}}{Q} \right )^4 ,
& Q/Q_{\rm c} \rightarrow \infty \\
\end{array} \right .
\end{eqnarray}
where we have used Eqs. (\ref{eq:qoqc}) and (\ref{eq:k11}),
\begin{eqnarray}
\chi_0 &\equiv& \mu_0 (2\pi/\phi_0)^2 \rho_{\rm s} d
\\ \nonumber
&=& 4\pi\alpha^2 \epsilon_r \frac{d}{\ell}
\left(\frac{\rho_s}{e^2/4\pi\epsilon\ell}\right)
\sim 3 \times 10^{-7} ,
\end{eqnarray}
and the numerical estimate of $\chi_0$ is given for the
``typical'' GaAs sample described in Sec.~\ref{sec:singsols}.
Here $\alpha\approx 1/137$ is the fine-structure constant.
We have plotted the susceptibility in Fig.~\ref{hannafig7}.
Note that near the CI transition, the susceptibility
diverges like $1/(Q-Q_{\rm c})$, with logarithmic corrections.

It might be possible to measure the SL magnetization or even
the SL magnetic susceptibility by varying the gate voltages of the sample,
in order to adjust $Q_c$ and tune close to the CI transition.
Measuring $\chi_{\rm SL}$ might be possible using AC modulation
of the gate voltages in order to AC modulate the layer imbalance $m_z$
and therefore the critical wave vector $Q_c$.  By such a method, the
ratio $Q/Q_c$ could be AC modulated just above and below unity,
allowing $\chi_{\rm SL}$ to be determined at or near the CI transition.
As an example, we compute here that part of
$\partial M_{\rm SL}/\partial m_z$ which is proportional to $\chi_{\rm SL}$:
\begin{eqnarray}
\label{eq:dmdm}
\frac{\partial M_{\rm SL}}{\partial m_z} &\sim&
\frac{\partial Q_{\rm c}}{\partial m_z}
\frac{\partial M_{\rm SL}}{\partial Q_{\rm c}}
\sim -\frac{\partial Q_{\rm c}}{\partial m_z} \frac{Q}{Q_{\rm c}}
 \frac{\partial M_{\rm SL}}{\partial Q}
\\ \nonumber
&=& -\frac{1}{\mu_0} \frac{\phi_0}{2\pi d}
\frac{\partial Q_{\rm c}}{\partial m_z} \frac{Q}{Q_{\rm c}} \chi_{\rm SL}
\\ \nonumber
&=& \frac{m_z}{1-m_z^2} \frac{M_0}{2} \frac{Q}{Q_{\rm c}}
\left( \frac{\rho_{\rm s}}{K_{11}} - 1 \right) ,
\end{eqnarray}
where we have used Eq~(\ref{eq:qcmz}) in the last line.
Equation (\ref{eq:dmdm}) shows that $\partial M_{\rm SL}/\partial m_z$
has a contribution proportional to $\chi_{\rm SL}$, which diverges
like $\rho_{\rm s}/K_{11}$ near the CI transition.
However, Eq.~(\ref{eq:dmdm}) vanishes for balanced layers ($m_z=0$);
thus, layer imbalance is required to measure $\partial M_{\rm SL}/\partial m_z$.
It turns out that when the layers are not balanced ($m_z\ne 0$),
the solitons have a nonzero electric dipole moment per length;\cite{rippled}
this changes the intersoliton repulsion from being exponentially weak
to a power law.
The net result is that the stiffness $K_{11}$ near the CI transition
is strengthened from being essentially linear in $(Q-Q_{\rm c})$ to
being proportional to $\sqrt{Q-Q_{\rm c}}$, so that the divergence in
$\chi_{\rm SL}$ becomes an inverse square-root singularity.
In real samples, we expect this singularity to be smoothed out
by finite temperature and disorder.

In practice, $m_z$ is varied by adjusting gate voltages.
For a small change $\delta V_G$ in a gate voltage away from
balance ($m_z=0$), the change
$\delta m_z$ in the layer imbalance is linearly proportional to $\delta V_G$.
For the ``typical'' GaAs sample described in Sec.~\ref{sec:singsols},
a rough estimate indicates that
\begin{equation}
\delta m_z \sim \frac{2\delta V_G}{e D_G n_T/(\epsilon_r\epsilon_0)}
\sim 3~\delta V_G/{\rm volt} ,
\end{equation}
if we take $D_G=10^{-6}$~m to be the distance
between the gate and the double layers.
There is an additional complication in changing the layer imbalance
by adjusting one of the gate voltages.
Unless both the back and front gates are adjusted together in a coordinated
way, the change in gate voltage $\delta V_G$ will also change the total
filling factor $\nu_T$ by an amount $\delta\nu_T$, where
\begin{equation}
\delta \nu_T \sim \frac{\delta V_G}{e D_G n_T/(\epsilon_r\epsilon_0)}
\sim 1.5~\delta V_G/{\rm volt} ,
\end{equation}
if $D_G=10^{-6}$~m.

\section{Summary}

In the presence of a sufficiently strong parallel magnetic field,
a $\nu_{\rm T}=1$ 2LQH device undergoes a transition to a
soliton-lattice (SL) state.
We have investigated the ground-state properties of the SL state
for all values of the parallel magnetic field,
with an eye towards possible experimentally measurable effects.

We found that the SL contribution to the orbital magnetization
rises in the commensurate phase ($Q<Q_c$) with $Q$, and quickly
drops to zero in the incommensurate phase ($Q>Q_c$).
An estimate of the size of the SL magnetization shows that it could
be detected by sensitive magnetometry
techniques.\cite{tormag,squid,cantilev}
The SL magnetic susceptibility shows a singularity at $Q=Q_c$, and
it was proposed that this signature of the CI transition might be
detected by varying the gate voltages so as tune close to the CI transition.

The longitudinal and transverse SL stiffnesses were computed
and used to estimate the temperature of the KT transition,
which might be indicated experimentally by an increase
in $\rho_{xx}$ at the transition.
A more sensitive signal of the KT transition would be obtained by
measuring the transresistivity (Coulomb drag) as a function of temperature,
although this would require separately contactable layers.
However, the leakage currents produced by any sizeable interlayer tunneling
might make it impractical to set up oppositely directed currents in each layer.

In this paper, we have largely neglected the existence of disorder.
As pointed out by Fisher\cite{mpaf} and discussed by Read\cite{read},
randomness in the tunneling $t$ due, for example, to small variations
in the barrier thickness, pins the SL domain walls randomly and
destroys the long-range order, no matter how weak the randomness.
This puts limits on how closely $Q$ can approach $Q_{\rm c}$
-- i.e., how small $\epsilon$ can be.

We have not discussed here the dynamics of the solitons -- i.e.,
the motion of individual solitons\cite{kyri} or the collective motion
of the SL\cite{cote} -- which also produce experimental signatures
of the incommensurate SL state.
We have focused instead on ground-state properties.
One of us has recently found that when the layer densities are made
unequal by adjusting the gate voltages of the device (keeping the
total filling factor equal to one), the layer densities in the SL
state become ``rippled'', resulting in a dipole density wave.\cite{rippled}
Preliminary calculations show that the sudden onset of such
a rippled SL state may give large contributions to the
differential capacitance of 2LQH systems, especially for the
interlayer capacitance (``Eisenstein ratio'').\cite{jpe,jungwirth}
Further work on the properties of the rippled state is in progress,
but it may be that sensitive measurements of differential capacitance
in unbalanced ($\nu_1\ne\nu_2$) 2LQH systems
could detect the CI transition.

\section{Acknowledgments}

C.B.H. would like to thank B. I. Halperin, N. Read, and T. C. Lubensky
for helpful discussions,
and the Institute for Theoretical Physics (ITP) at the University of
California, Santa Barbara, where part of this work was carried out,
for providing support through the ITP Scholars Program.
This work was supported by an award from Research Corporation
and by the National Science Foundation under grants DMR-9972332,
DMR-9714055, and DMR-0087133.

\addcontentsline{toc}{part}{Figure Captions}

\begin{references}
\bibitem{bakcorrect}
Equation~(\ref{eq:qsofq}) for $\bar{Q}_{\rm s}/Q$
and Eq.~(\ref{eq:qoqc}) for $Q/Q_{\rm c}$ in this work
are equivalent to Eqs. (2.7) and (2.8), respectively,
in Ref. \onlinecite{bak}, after correcting for misprints.
\end{references}
\end{document}